# The physics of diamagnetic levitation

A. Schilling

Dept. of Physics, University of Zurich, Winterthurerstrasse 190, CH-8057 Zürich

**Abstract**

We formulate a problem on diamagnetic levitation, which may be suitable for specialized high-school or first-year students in physical sciences. We guide the students, step-by-step, through the physics of diamagnetic levitation. The calculations are simplified by assuming a ring-shaped geometry of the diamagnetic object residing above a magnetic dipole. This problem was originally intended for the International Physics Olympiad 2016 (IPHO 2016), but was finally deemed surplus and therefore not set.

# The physics of diamagnetic levitation

A. Schilling
Dept. of Physics, University of Zurich, Winterthurerstrasse 190, CH-8057 Zürich

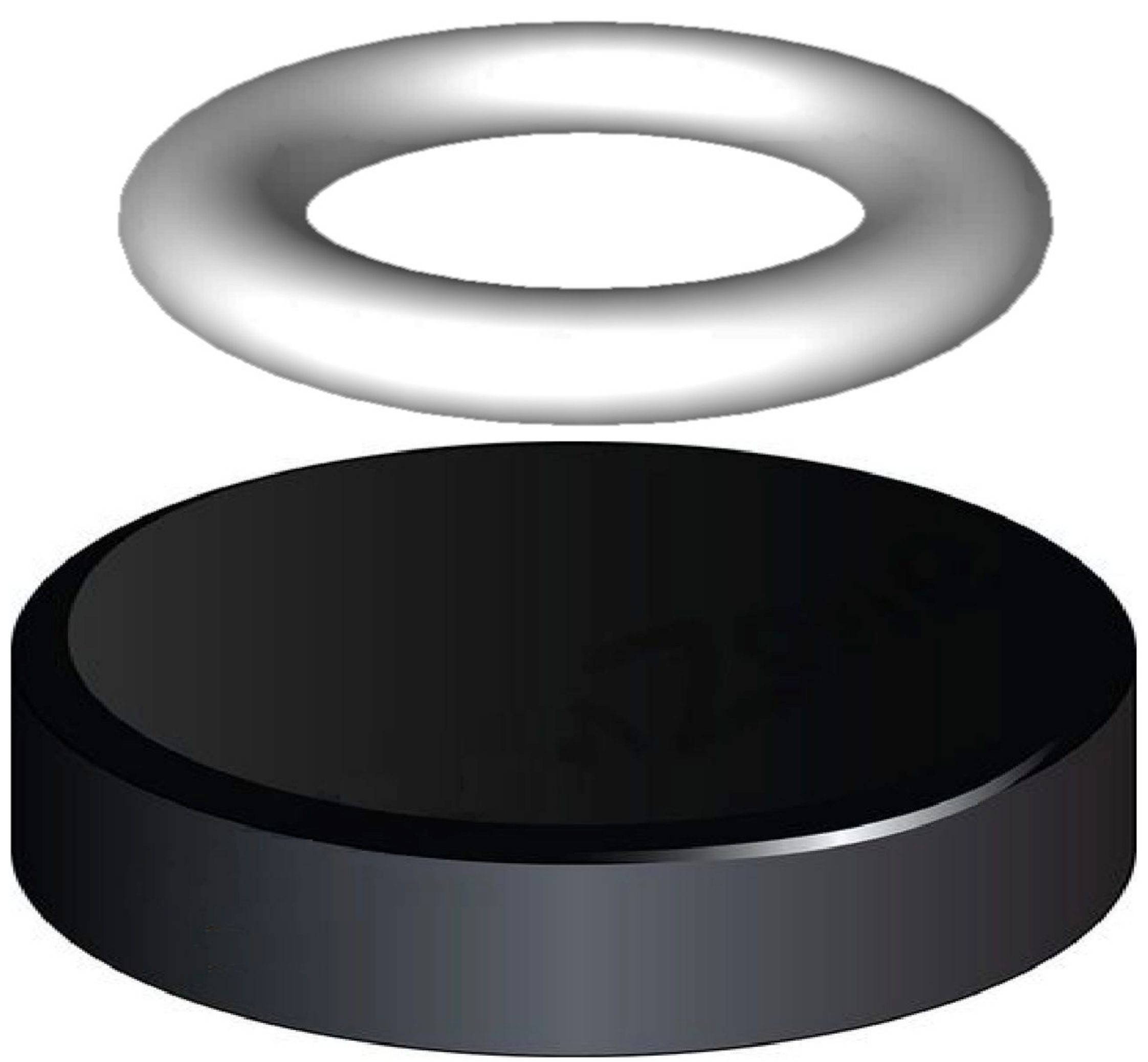

## The physics of diamagnetic levitation

You may have seen the superconducting train levitating on permanent magnets here on our campus, which allows for a motion that is almost free of friction (see Figure 1 and [1]). The family of superconductors used in this experiment, known as "high-temperature superconductors", was originally discovered by Karl Alex Müller at our University and his colleague Georg Bednorz at IBM Rüschlikon, Switzerland [2]. The current record for the highest transition temperature to superconductivity at ambient pressure (as of January 2021) is still in a compound first reported by the author of this article [3], meanwhile optimized to 138 K (≈ -135 ºC)) [4].

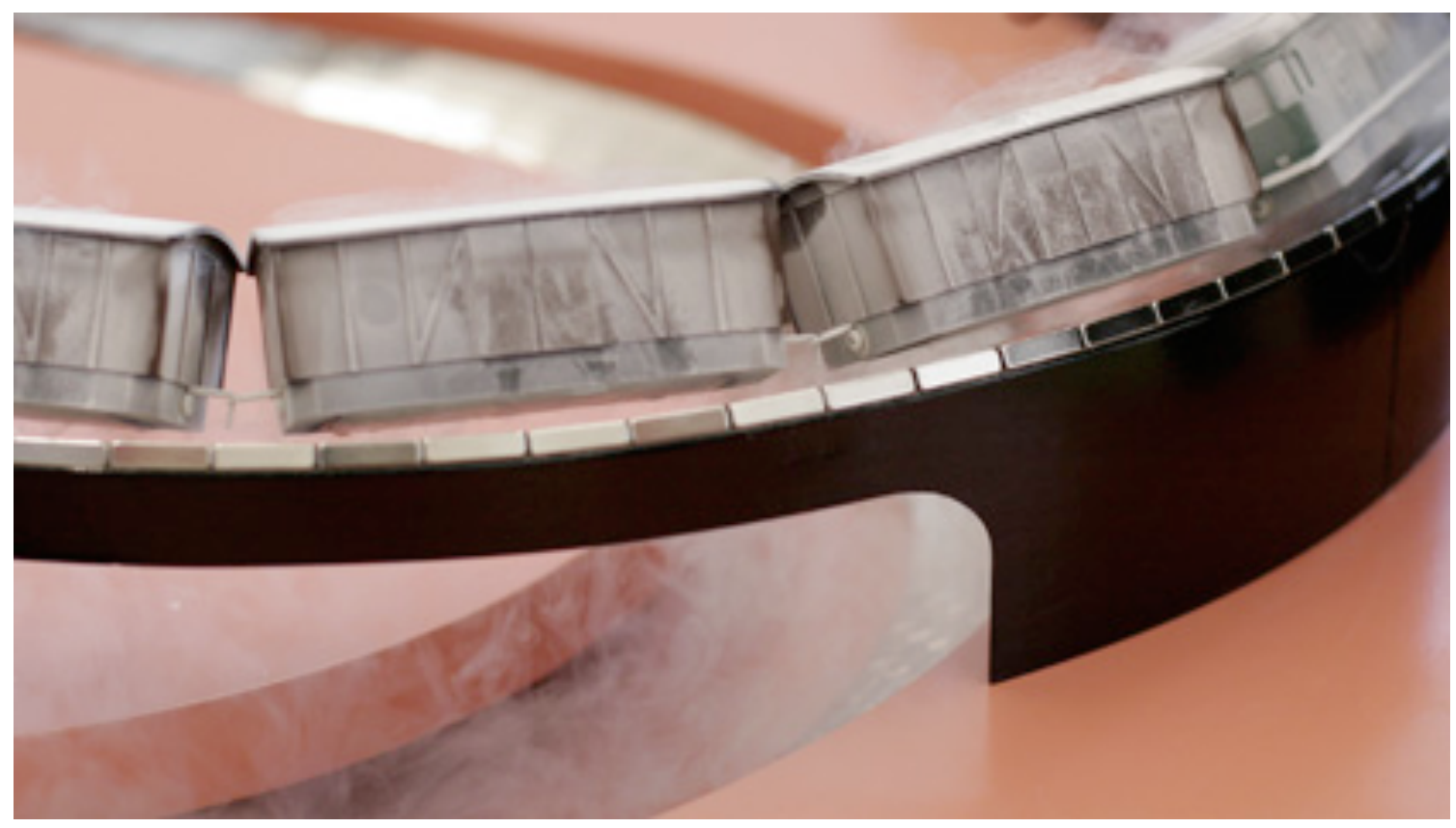

*Figure 1: Train levitated by superconductors on a magnetic trail*

Superconductors react to an external magnetic field $B_{ex}$ in such a way that they tend to expel it completely from their interior: a superconductor is a *perfect diamagnet*. This means that when the superconductor is exposed to an external magnetic field $B_{ext}$, it creates locally a perfectly "compensating field" $\mu_0 M = -B_{ext}$ ($M$ is called a "magnetization", with $\mu_0 = 4\pi \times 10^{-7}$ Tm/A) so that the sum of $\mu_0 M$ and the external magnetic field inside the superconductor is zero ($M$ has the units A/m, and $B_{ext}$ the unit Tesla).

In terms of a material constant, we can ascribe to a superconductor a *dimensionless magnetic susceptibility* $\chi = \mu_0 M / B_{ext}$, which represents the ratio of the "compensating field" and the external magnetic field. From what we have said above, a perfect diamagnet has $\chi = \mu_0 M / B_{ext} = -1$. You will learn below that diamagnetic materials (i.e., materials with $\chi < 0$) can levitate above permanent magnets under certain circumstances.

Unfortunately, superconductors at normal pressure have to be cooled way below room temperature to operate. There are other materials with diamagnetic properties even at room temperature, but their dimensionless magnetic susceptibilities are usually only a tiny fraction of $\chi = -1$.

One of these materials is pyrolytic graphite. A piece of pyrolytic graphite can levitate in the presence of strong enough and suitably arranged permanent magnets even at room temperature, as you can see it on the picture below (Figure 2).

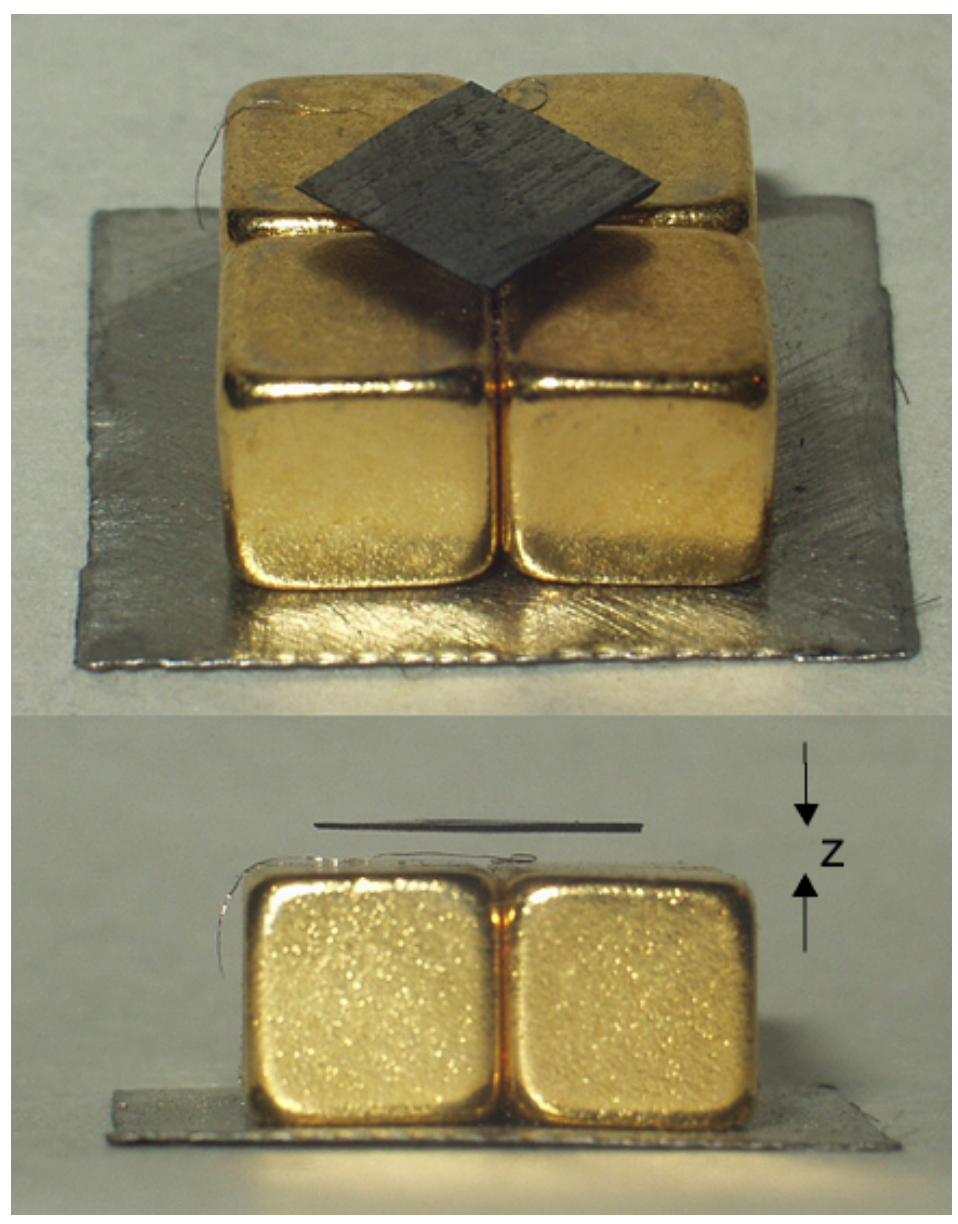

*Figure 2: A sheet of pyrolytic graphite levitating on an array of permanent magnets*

In Figure 2, four cube shaped permanent magnets are placed is such away that a stable levitation is possible. To achieve this, the poles of the cubes are alternating (north-south-north-south).

The correct quantitative analysis of levitation in this geometry is rather complicated. A much easier geometry that we want to consider here is the levitation of a diamagnetic torus above a cylindrical permanent magnet (see Figure 3, left).

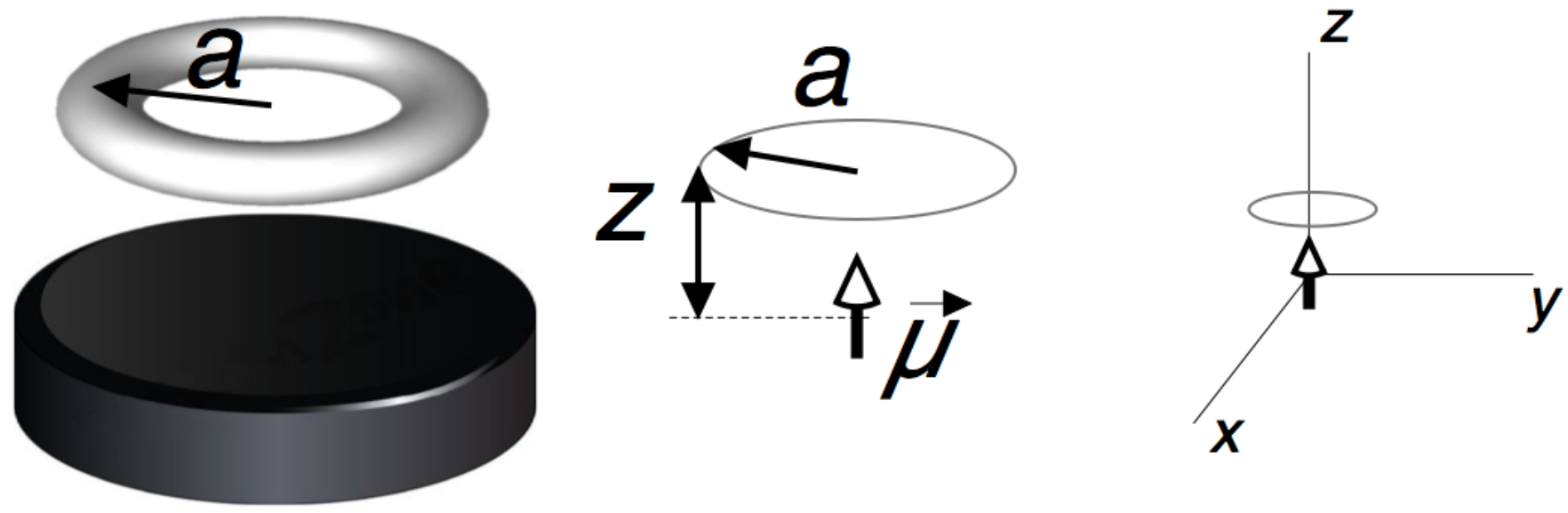

*Figure 3, left: a diamagnetic torus with radius a levitates above a cylindrical magnet; middle: an idealized situation with a diamagnetic circle above a magnetic dipole moment $\mu$; right: choice of the coordinate system*

**Approximations: We approximate the cylindrical permanent magnet by a point-like magnetic dipole "$\vec{\mu}$" (Figure 3, middle), and the torus by a circle with radius *a* and infinitely small cross-section *A*, so that the total mass *m* is equally distributed along the circumference of the circle.**

In the following, we guide you through several considerations and calculations, which should enable you to study this levitation problem and investigate under which circumstances a stable levitation is possible.

### 1) The magnetic dipole moment $\vec{\mu}_{mag}$ of the permanent magnet

In general, the magnetic dipole moment $\vec{\mu}$ of a magnetized volume $V$ (here we choose the volume $V$ of the cylindrical magnet) is a vector, given by $\vec{\mu} = V\vec{M}$. It is thought to point from the south to the north pole of the considered magnet. We will now estimate the value of the magnetic dipole moment $\mu_{mag} = |\vec{\mu}_{mag}|$ of the permanent magnet (to distinguish the magnetic dipole moment produced by the permanent magnet from other magnetic moments in this problem, we call it "$\vec{\mu}_{mag}$"). The strongest available permanent magnets consist of an alloy of Nd, Fe and B, and if they are fully saturated, they produce in their interior a magnetization $\mu_0 M_{mag} \approx 1.3$ T. The cylindrical permanent that we consider here has a radius $R = 2$ mm and a height $h = 2$ mm.

**Question a)** Please calculate the absolute value of the magnetic dipole moment $\mu_{mag}$ of this magnet, expressed by $M_{mag}$, $R$ and $h$, and as a number with correct units.

### 2) The magnetic field $\vec{B}(\vec{r})$ created by the magnetic dipole moment

Now we consider $\vec{\mu}_{mag}$ as a known quantity, and focus on the magnetic field created by it. We choose the origin of our coordinate system at the position of the magnetic dipole (Figure 3, right). It is known that the magnetic field $\vec{B}(\vec{r})$ at the position $\vec{r}$ produced by a magnetic dipole moment $\vec{\mu}_{mag}$ is

$$\vec{B}(\vec{r}) = \mu_0 \frac{3\vec{r}(\vec{\mu}_{mag} \circ \vec{r}) - \vec{\mu}_{mag} r^2}{4\pi r^5} .$$

The diamagnetic ring with radius $a$ is thought do reside above the dipole in a distance $z$ in the direction of $\vec{\mu}_{mag}$, with its center directly above the dipole so that the symmetry axis of circle is parallel to $\vec{\mu}_{mag}$ (Figure 3, middle).

**Question b)** Please calculate all components of the magnetic field vector $\vec{B}(\vec{r})$ at the position $x = a$ and $y = 0$ on the ring, expressed by $\mu_{mag}$, $a$ and $z$.

### 3) Levitation

We now turn to the physics of levitation. In equilibrium, the forces acting on the diamagnetic ring determine in which distance $z$ it is levitating above the magnet.
We first consider a line element d$r$ of the circle at the position $x = 0$ (see Figure 4).

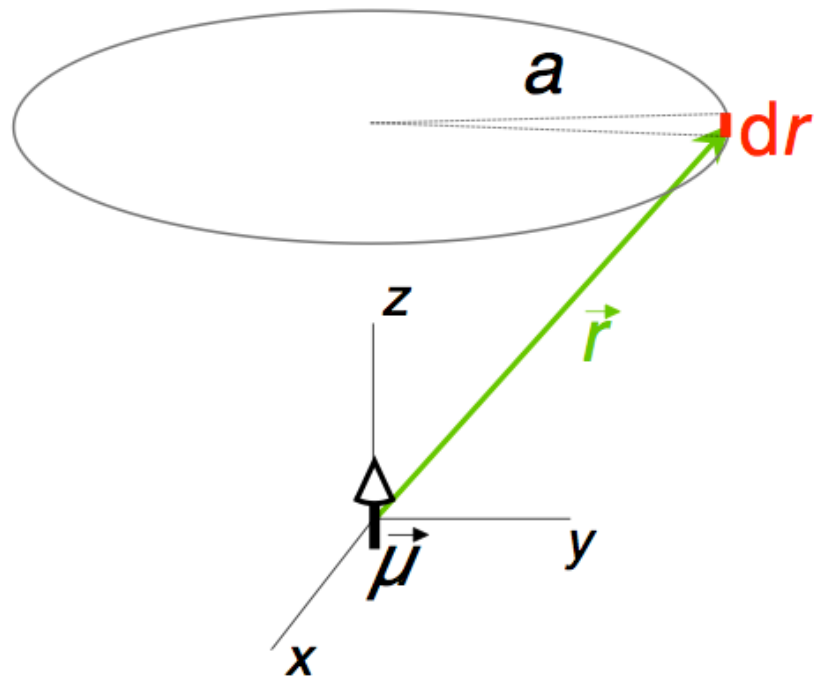

*Figure 4: Line element dr at the position x = 0 on the diamagnetic ring*

Below you will calculate the magnetic potential energy of this line element (we call it "partial " magnetic potential energy), then that of the whole ring, and finally you can derive from it the magnetic force acting on the ring.

As you will see, the partial magnetic potential energy $dE_{pot}$ of this line element d*r* depends on the partial magnetic dipole moment $d\vec{\mu}_{gr} = dV\vec{M}_{gr}$ of the line element with the volume d*V* filled with pyrolytic graphite (therefore we use the index "*gr*"), and on $\vec{B}(\vec{r})$ created by the magnetic dipole at the position $\vec{r}$ of d*r* (see Figure 4).

With the given magnetic susceptibility $\chi_{gr}$ of pyrolytic graphite and $\vec{B}(\vec{r})$, you can express $\vec{M}_{gr}$ and $d\vec{\mu}_{gr} = dV\vec{M}_{gr}$ in a straightforward way.

### 3.1) Magnetic potential energy of the ring

You may have learnt or heard that the potential energy stored by a magnetic moment $\vec{\mu}$ in the presence of a magnetic field $\vec{B}$ is given by the scalar product $E_{pot} = -\vec{\mu} \circ \vec{B}$. In our problem, the partial magnetic potential energy $dE_{pot}$ of the line element d*r* with partial magnetic moment $d\vec{\mu}_{gr}$ is $dE_{pot} = -d\vec{\mu}_{gr} \circ \vec{B}$. The total magnetic potential energy $E_{ring}$ of the ring is then simply the sum of all partial magnetic potential energies, summed over all line elements d*r* of the ring.

**Question c)** Show step by step, that the total magnetic potential energy $E_{ring}$ of the ring, expressed by $\chi_{gr}$, $\mu_{mag}$, $a$, $z$ and the cross section $A$ of the torus, is

$$E_{ring} = \mu_0 \frac{-\chi_{gr} a A \mu_{mag}^2}{8\pi} \frac{\left(a^2 + 4z^2\right)}{\left(a^2 + z^2\right)^4}.$$

### 3.2) Force derived from the magnetic potential energy

In the following you can use the expression for $E_{ring}$ given in question c). The *z*-component of the resulting force can be derived from it using

$$F_z = -\frac{\mathrm{d}E_{ring}}{\mathrm{d}z}.$$

**Question d)** So far we have ignored force components in directions other than $z$. What can you tell about the *total* force on the whole ring in the $x$ and the $y$ directions?

**Question e)** Please calculate the $z$-component of the magnetic force on the whole ring, expressed by $\chi_{gr}$, $\mu_{mag}$, $a$, $z$ and the cross section $A$ of the torus.

### 4) Levitation condition

Knowing the $z$-component of the magnetic force on the whole ring, you can now find an equation for the levitation height $z$ of the ring above the magnetic dipole. We assume that besides $\chi_{gr}$ and $\mu_{mag}$, the mass density $\rho$ of pyrolytic graphite and the gravitational acceleration g are also known.

**Question f)** Find an expression that describes how the ring radius $a$ depends on levitation height $z$, if the ring levitates in equilibrium of forces. Here, $\chi_{gr}$, $\mu_{mag}$, $z$, g and $\rho$ are given quantities. Please also sketch the function $a(z)$ using the numeric values of $\mu_{mag}$, $\chi_{gr} = -450\times10^{-6}$, $\rho = 2100$ kg/m$^3$ and g = 9.81 m/s$^2$ over the full range of possible $z$ values. If you were not successful to calculate $\mu_{mag}$ in a), use here and in the following the numerical value $\mu_{mag} = 0.022$ in correct SI units.

**Question g)** Which is the theoretically maximum achievable levitation height $z_{max}$, and for which $a$ would it be realized? $\chi_{gr}$, $\mu_{mag}$, g and $\rho$ are given quantities.

### 5) Maximum possible ring radius

**Question h)** Which is a maximum possible ring radius $a$ for which magnetic and gravitational forces can cancel each other? $\chi_{gr}$, $\mu_{mag}$, g and $\rho$ are given quantities.

### 6) Stability with respect to *z*-direction

Assume that the ring is at first slightly vertically displaced below its levitation height. If released, it will either move back upwards to the original position (i.e., that position was stable), or the ring will drop down (i.e., the position was unstable).

**Question i)** Develop a criterion for a stable levitation with respect to the $z$ direction: for which values of $z$ is the levitation stable? $\chi_{gr}$, $\mu_{mag}$, g and $\rho$ are given quantities.

## 7) Stability with respect to radial direction

Now we assume that the ring is slightly *horizontally* displaced at its levitation height. If released, it will either move back to the original position (i.e., that position was stable) or the ring will be sidewise expelled from it (i.e., the position was unstable).

**Question k)** Develop a criterion for a stable levitation with respect to the *radial* direction: for which values of $z$ is the levitation stable? Again, $\chi_{gr}$, $\mu_{mag}$, g and $\rho$ are given quantities.

Summarize the results from questions i) and k) to identify the $z$ values for which a levitation is stable with respect to disturbances in both $z$ and radial directions.

## 8) Oscillation frequency

If the ring is levitating in a stable position and subsequently slightly displaced from this position by a small displacement $\Delta z$ in the $z$ direction, there will be a restoring force $\Delta F_z$. The ratio of this restoring force and the small displacement $\Delta z$ around the equilibrium position defines a "spring constant" $k = -\Delta F_z / \Delta z$, similar to the case of an extended spring. If you were successful in answering question i), it is easy to find an expression for this spring constant.

**Question l)** Calculate the "spring constant" $k$, where $a$, $z$, $\chi_{gr}$, $\mu_{mag}$, g and $\rho$ are given quantities, for the limit $\Delta z \to 0$.

If the ring is released after such a small displacement in the $z$ direction, it will oscillate around its equilibrium position. The resulting oscillation frequency is given by $f = \sqrt{k/m} / 2\pi$.

**Question m)** Calculate the oscillation frequency from the known quantities $\chi_{gr}$, $\mu_{mag}$, g and $\rho$ for the maximum and the minimum stable levitation heights.

## 9) What changes if a superconducting material is used?

We have seen in the introduction that ideal superconductors have a dimensionless magnetic susceptibility $\chi = -1$. We now assume that we have a superconducting ring instead of ring of pyrolytic graphite.

**Question n)** What would be the maximum stable levitation height $z_{\max,stable}$ for a high-temperature superconductor with $\rho = 6400$ kg/m$^3$ if $\chi = -1$, and if the same magnet is used as in the previous examples?

**SOLUTIONS:**

Solution a): $\mu_{mag} = \pi R^2 h M_{mag} = 0.0260\ \mathrm{Am^2}$

Solution b): $$\vec{B} = \frac{\mu_0 \mu_{mag}}{4\pi (a^2 + z^2)^{5/2}} \begin{pmatrix} 3az \\ 0 \\ 2z^2 - a^2 \end{pmatrix}$$

Solution c):

As indicated in the problem, $\mathrm{d}E_{pot} = -\mathrm{d}\vec{\mu}_{gr} \circ \vec{B}$ for the line element $\mathrm{d}r$, with $\mathrm{d}\vec{\mu}_{gr} = \mathrm{d}V\vec{M}_{gr} = A\vec{M}_{gr}dr$ and $\mu_0 \vec{M}_{gr} = \chi_{gr}\vec{B}$. Therefore,

$$\mathrm{d}E_{pot} = -\mathrm{d}\vec{\mu}_{gr} \circ \vec{B} = A\vec{M}_{gr} \circ \vec{B}dr = -A\chi_{gr}\vec{B}^2 dr / \mu_0$$

For symmetry reasons, $\vec{B}^2$ is the same on the whole ring, and therefore

$$E_{ring} = -A\chi_{gr}\vec{B}^2 2\pi a / \mu_0 . \qquad \text{Note that } -\chi_{gr} > 0\,!$$

From b), we have $\vec{B}^2 = \dfrac{\mu_0^2 \mu_{mag}^2}{16\pi^2 (a^2+z^2)^5}(5a^2z^2 + 4z^4 + a^4) = \dfrac{\mu_0^2 \mu_{mag}^2 (a^2 + 4z^2)}{16\pi^2 (a^2+z^2)^4}$.

and $$E_{ring} = -A\chi_{gr}\vec{B}^2 2\pi a / \mu_0 = -\mu_0 \frac{\chi_{gr} a A \mu_{mag}^2}{8\pi} \frac{(a^2+4z^2)}{(a^2+z^2)^4}.$$

Solution d): they are zero for symmetry reasons, or because

$$\mathrm{d}E_{ring} / \mathrm{d}x = \mathrm{d}E_{ring} / \mathrm{d}y = 0$$

Solution e): $$F_z = -\mu_0 \frac{3\chi_{gr} a A \mu_{mag}^2}{\pi} \frac{z^3}{\left(a^2+z^2\right)^5} > 0 \quad (-\chi_{gr} > 0)$$

Solution f): Set $$-\mu_0 \frac{3\chi_{gr}\mu_{mag}^2}{2\pi^2} \frac{z^3}{\left(a^2+z^2\right)^5} = \rho g$$

$$a^2 = \sqrt[5]{-\mu_0 \frac{3\chi_{gr}\mu_{mag}^2}{2\pi^2 \rho g} z^3} - z^2 = \sqrt[5]{\beta^7 z^3} - z^2 \text{, with } \beta \equiv \left(-\mu_0 \frac{3\chi_{gr}\mu_{mag}^2}{2\pi^2 \rho g}\right)^{1/7} = 3.11\mathrm{mm}$$

Sketch of $a(z)$ see below

Solution g): from the above result $a(z)$ in the limit a -> 0,

$$z_{\max} = \left( -\mu_0 \frac{3\chi_{gr}\mu_{mag}^2}{2\pi^2 \rho g} \right)^{1/7} = \beta$$

$$z_{\max} \approx 3.11\text{mm}$$

Solution h): from the above result for $a(z)$, set $d(a^2)/dz = 0$, which yields with

$$z_{a\max} = (3/10)^{5/7}\beta \qquad a_{\max} = z_{a\max}\sqrt{7/3} = 0.646\beta$$

$z_{a\max}$ = 1.32 mm, $a_{\max}$ = 2.01 mm.

Solution i): $d^2E_{ring}/dz^2$ must be > 0 (or $dF_z/dz$ < 0).

$$\frac{dF_z}{dz} = -\mu_0 \frac{3\chi_{gr} aA\mu_{mag}^2}{\pi} \frac{3a^2z^2 - 7z^4}{(a^2+z^2)^6} \qquad \text{(positive sign, because } -\chi_{gr} > 0\text{).}$$

$$\frac{dF_z}{dz} < 0 \text{ for } z > \sqrt{3/7}a = 0.655\ a.$$

Surprisingly, this result does depend neither on the properties of the magnet nor of those of pyrolytic graphite! To obtain numbers, inserting $z = 0.655\ a$ in $a(z)$ gives

$$z > (3/10)^{5/7}\beta = z_{a\max} = 1.32 \text{ mm.}$$

Solution k): consider only one line element $dr$ of the ring. Every horizontal displacement of the whole ring results in a certain radial and a tangential displacement component of this line element. The tangential component does not change $dE_{ring} = drE_{ring}/2\pi a$ of the line element due to the radial symmetry. Only the radial displacement of the line element is relevant here. Therefore, $d^2(E_{ring}/2\pi a)da^2$ must be > 0 (here, $a$ replaces the variable $r$).

$$\frac{d^2}{da^2}\left(\frac{E_{ring}}{2\pi a}\right) \propto \frac{(7a^4 + 42a^2z^2 - 5z^4)}{(a^2+z^2)^6} > 0 \text{ for } z < \sqrt{\frac{42+\sqrt{1904}}{10}}a = 2.926\ a.$$

This limit is valid for all line elements showing a horizontal displacement, and is therefore valid for the whole ring. To obtain numbers, inserting $z = 2.926\ a$ in $a(z)$ gives

$$z < 0.924\beta = 2.88 \text{ mm.}$$

Summary: stable levitation for $(3/10)^{5/7}\beta < z < 0.924\beta$, **1.32 mm < $z$ < 2.88 mm**.

Overview of the solutions of questions f-k:

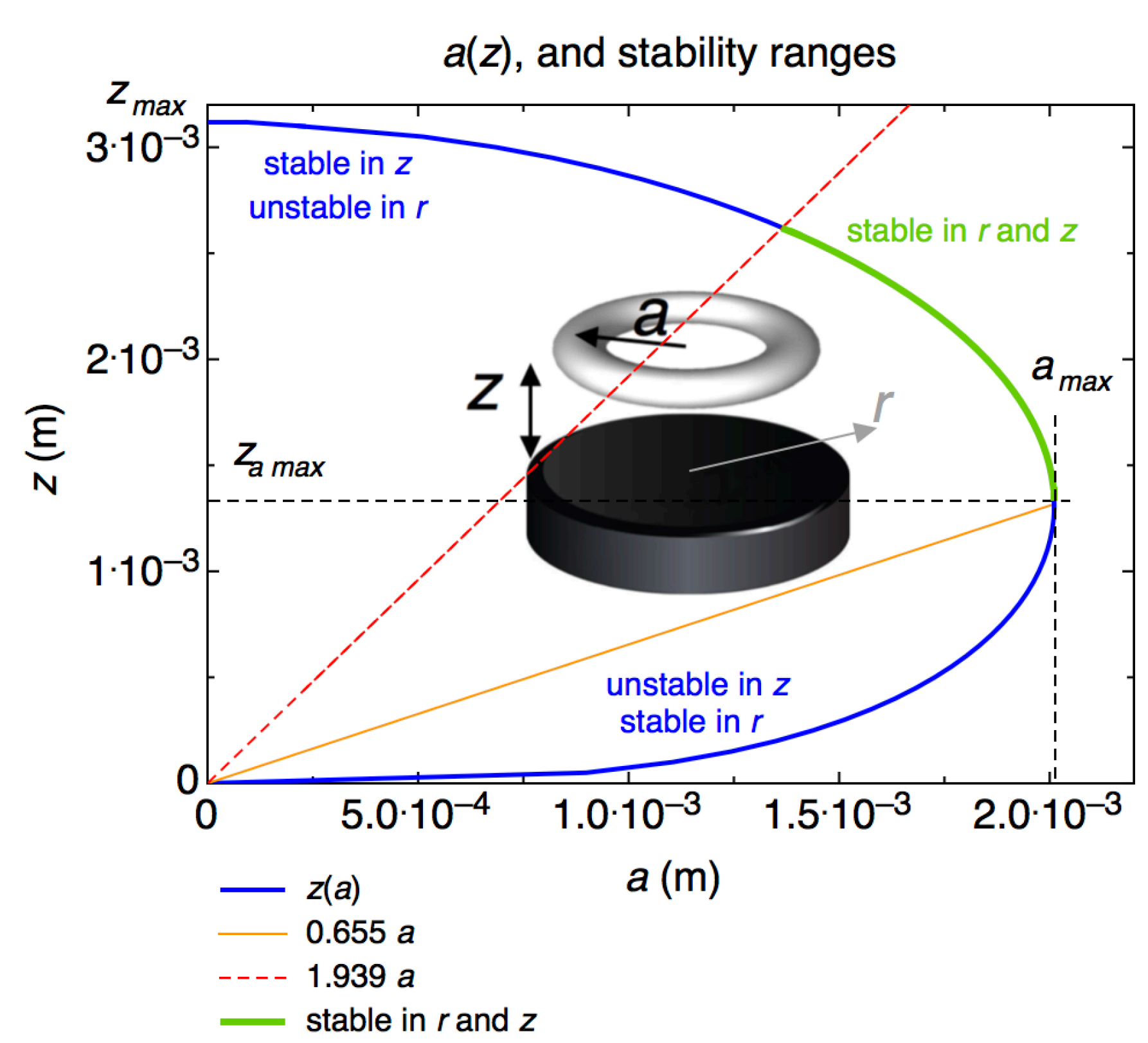

Solution l): The spring constant is, using the solution of question i),

$$k = -\frac{\mathrm{d}F_z}{\mathrm{d}z} = \mu_0 \frac{3\chi_{gr} a A \mu_{mag}^2}{\pi} \frac{3a^2z^2 - 7z^4}{(a^2+z^2)^6} = -2\pi\rho g a A \beta^7 \frac{3a^2z^2 - 7z^4}{(a^2+z^2)^6}$$

Solution m): With the mass of the ring, $m = 2\pi a A \rho$, $f = \sqrt{g\beta^7 \frac{7z^4 - 3a^2z^2}{(a^2+z^2)^6}} / 2\pi$.

From the solution of question f), $a^2 = \sqrt[5]{\beta^7 z^3} - z^2$, where $\beta$ is defined above, and

$$f = \sqrt{g\left(10 z^{2/5} \beta^{-7/5} - \frac{3}{z}\right)} / 2\pi .$$

From question k) we know that the range of stable $z$-values is $(3/10)^{5/7}\beta < z < 0.924\beta$. At the minimum value with $(3/10)^{5/7}\beta = z$, $f \to 0$. At the maximum value, $z = 0.924\beta$, we have with $\beta \approx 3.11$ mm (question f) $f \approx 23$ Hz.

Solution n): From the solution k), we know that $z < 0.924\beta = z_{\max,stable}$.

Inserting the numbers in the definition of $\beta$ given in solution f), we have $\beta \approx 8.0$ mm and $z_{\max,stable} \approx 7.4$ mm.